# Observation of an Efimov spectrum in an atomic system

M. Zaccanti[*,1], B. Deissler[1], C. D'Errico[1], M. Fattori[1,2], M. Jona-Lasinio[1], S. Müller[3], G. Roati[1], M. Inguscio[1], & G. Modugno[1]

1 LENS and Physics Department, Università di Firenze, and INFM-CNR, Via Nello Carrara 1, 50019 Sesto Fiorentino, Italy
2 Museo Storico della Fisica e Centro Studi e Ricerche "E. Fermi", Roma, Italy
3 5. Physikalisches Institut, Universität Stuttgart, 70550 Stuttgart, Germany
* e-mail: zaccanti@lens.unifi.it

In 1970 V. Efimov predicted a puzzling quantum-mechanical effect that is still of great interest today. He found that three particles subjected to a resonant pairwise interaction can join into an infinite number of loosely bound states even though each particle pair cannot bind. Interestingly, the properties of these aggregates, such as the peculiar geometric scaling of their energy spectrum, are universal, i.e. independent of the microscopic details of their components. Despite an extensive search in many different physical systems, including atoms, molecules and nuclei, the characteristic spectrum of Efimov trimer states still eludes observation. Here we report on the discovery of two bound trimer states of potassium atoms very close to the Efimov scenario, which we reveal by studying three-particle collisions in an ultracold gas. Our observation provides the first evidence of an Efimov spectrum and allows a direct test of its scaling behaviour, shedding new light onto the physics of few-body systems.

From nuclei, atoms and molecules up to galaxies, our complex world is made up of many kinds of aggregates whose properties depend on the details of the interactions between their components. This scenario is expected to drastically change as one moves to the world of few neutral quantum particles. The physics of these systems is typically dominated by two-body interactions, which in the limit of vanishing collision energies can be described by a *single* parameter, namely the s-wave scattering length, independently from the nature of the particles and of the microscopic shape of their interaction[1,2]. If the two-body scattering length becomes resonantly large, the binding of few such particles into larger aggregates is predicted to become *universal*, in the sense that its properties depend only on the scattering length and few other global parameters[3].

These expectations have been so far verified only for two-body bound states[2], and even the seemingly simple case of three particles is still under investigation. In this frame, a landmark theoretical result was obtained in 1970 by Efimov[4,5]. He extended previous studies[6] to show that three identical bosons with large two-body scattering length $a$, even without two-body bound states, can form weakly bound trimer states with size greatly exceeding the characteristic range $r_0$ of the two-body potential. The binding properties of such states follow a universal behaviour, regardless of the microscopic peculiarities of their components and of their interaction. Efimov indeed identified an effective three-particle interaction potential of the form $-(s_0^2+1/4)/R^2$, where $R$ is the overall size of the three-body system and $s_0 \sim 1.00624$ is a universal parameter[4]. This simple potential is known to support an infinite number of bound states whose energy spectrum exhibits a peculiar geometric scaling where two consecutive states are linked by the relation $E_n = E_{n-1} \exp(-2\pi/s_0)$. This perfect scaling is predicted to apply only for the special case of a system with infinite $a$. For finite $a$, the number of bound states becomes of the order of $\ln(|a|/r_0)$, the scaling behaviour for the energies of the ladder is only approximate, but shows up in the relative values of $a$ at which the trimers dissociate into unbound particles. Analogous results, differing mainly because of a different value of the Efimov parameter $s_0$, have been predicted also for many combinations of three non-identical particles with at least two resonant interactions[7].

In view of its universality, the Efimov effect has sparked a strong interest in many fields of physics. However, despite an extensive search among different physical systems including nuclei, atoms and molecules[3,8-10], the peculiar Efimov spectrum has so far eluded observation. This is mainly due to the difficulty of realizing a system with resonant two-body interaction fulfilling the requirement $|a| \gg r_0$. Ultracold atomic gases have demonstrated to be a very appealing system in this respect, because of their ultra-low collision energies and because Feshbach resonances[11] allow to tune the scattering length over several orders of magnitude, offering the unprecedented possibility to reproduce in a single system the whole Efimov scenario. However, only individual Efimov states have been found so far through the study of atomic and molecular collisions in cesium atoms[9,10], and doubts about the realistic possibility of testing the scaling properties of the spectrum in experiments have been raised[12].

We here report on the discovery of the first two states of an Efimov spectrum in a system composed of ultracold potassium atoms with resonantly tunable two-body interaction. We find that the study of a single observable of the system, i.e. the recombination of three atoms into bound states, allows us to extract more information than expected by current theories about the Efimov states, and provides a test of their scaling behaviour. We find a scaling not far from the geometric universal one, despite the fact that our system is not expected to be fully in the universal regime. Our findings provide a test ground for advanced few-body theories, and open a new window for studying universal few-body phenomena.

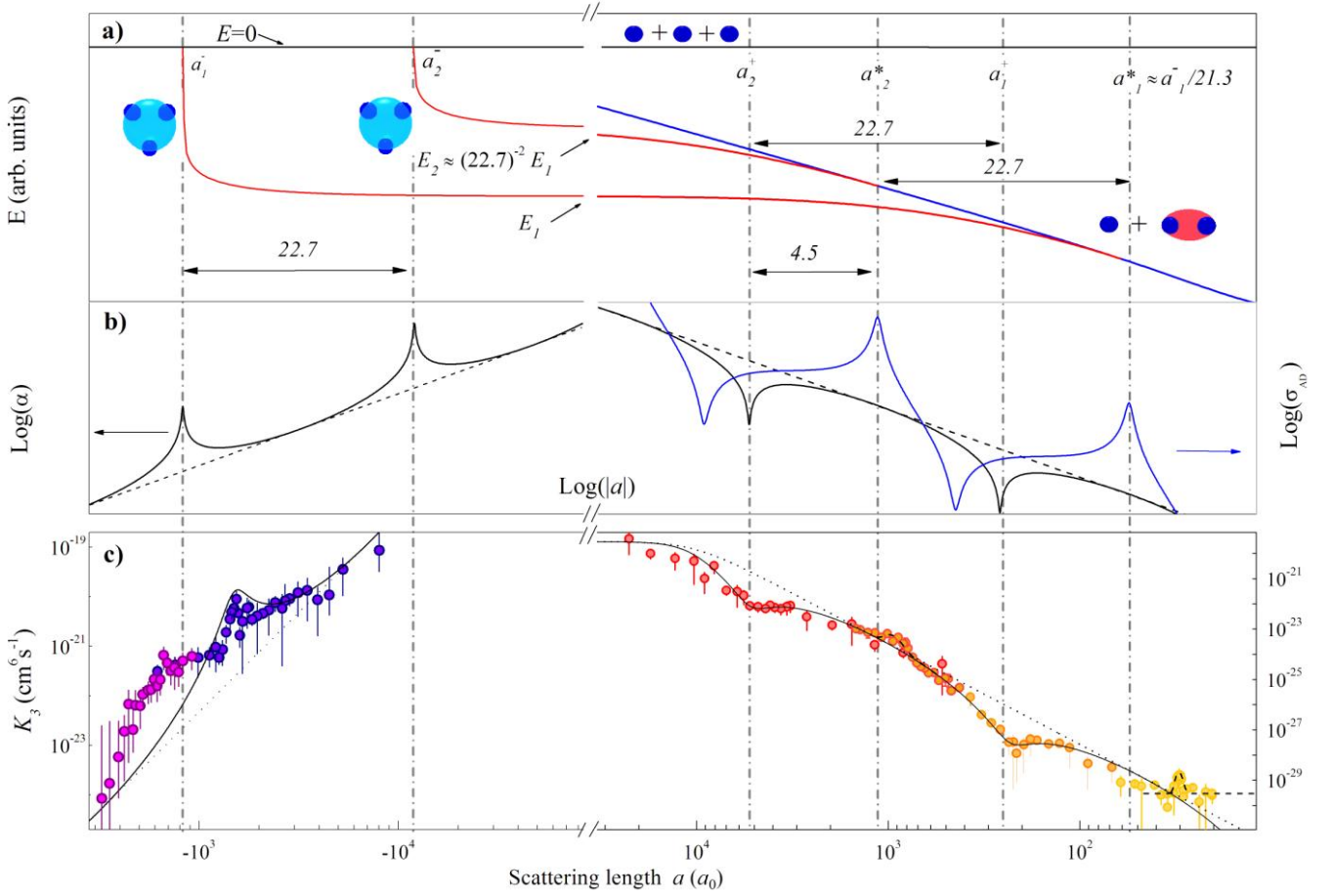

**Figure 1| Efimov spectrum. a,** Theoretical binding energy of two consecutive Efimov states (red) and of the dimer state (blue) in the universal regime versus the scattering length $a$. **b,** Theoretical three-body recombination rate $\alpha$ (black) and atom-dimer elastic cross section $\sigma_{AD}$ (blue). Vertical dash-dotted lines indicate the position of the detectable maxima and minima in the three-body observables, whose relevant scaling rules are summarized in Fig. 1a. **c,** Measured recombination coefficient $K_3$ in an ultracold potassium gas (circles), featuring deviations from the bare $a^4$ trend (dotted line), and fitted behaviour assuming a local universal trend for $K_3$ in the vicinity of the two recombination minima at $a>0$ and of the Efimov resonance at $a<0$ (continuous line), see text. The other two features due to the atom-dimer resonances $a^*_1$ and $a^*_2$, not expected by theory, are locally fitted with a gaussian profile superimposed to a constant background and to the universal behaviour, respectively (dashed lines). The various colours correspond to different data sets. For all data points, the error bars are the root sum squared of the s.e.m. resulting from the fit and of the uncertainty on the trap frequencies (see Methods).

## Geometric scaling of three-body observables

To understand the essence of our discovery, let us discuss at first the basic aspects of an Efimov spectrum for $|a|\gg r_0$. In this regime, the behaviour of the trimer states is universal, and the three-body observables can be described by simple analytic expressions, derived by a wide variety of theoretical methods (see Ref. 13 and references therein), often called *universal theory* (UT). The $a$-dependence of the energy of two consecutive Efimov trimers is sketched in Fig.1a, which shows also the weakly bound dimer state that exists for $a>0$, whose energy is $-\hbar^2/M a^2$, $M$ being the atomic mass. This picture shows how trimers dissociate into three free atoms for $a<0$ and into a dimer plus an atom for $a>0$ for decreasing $|a|$. How to detect the presence of the Efimov states in an atomic system? They are actually expected to affect three-body observables such as the rate $\alpha$ for recombination processes of the kind $b+b+b \rightarrow b_2^*+b$ and $b+b+b \rightarrow b_2+b$, and the cross-section for atom-dimer scattering events of the kind $b_2^*+b \rightarrow b_2^*+b$ and $b_2^*+b \rightarrow b_2+b$. Here $b_2^*$ and $b_2$ represent a weakly- and a deeply-bound dimer, respectively. These observables are predicted to display an oscillatory behaviour on a logarithmic scale[13-16] with period determined by the same universal parameter as $\exp(\pi/s_0) \sim 22.7$, superimposed to the trend expected in the absence of Efimov states[16-18] (see Fig 1b). Their investigation can therefore provide information about the geometric scaling of the energy spectrum. Note how recombination maxima arise as the

trimer states hit the atomic continuum at $a^-_n<0$, opening new decay channels towards deeply bound two-body states, which enhance the recombination rate. The recombination minima arising at specific $a^+_n>0$ can instead be explained as the destructive interference between two different decay pathways[19]. The resonances in the elastic and inelastic components of the atom-dimer cross-section appear instead each time an Efimov state crosses the atom-dimer threshold at $a^*_n>0$. In the regime $r_0<<|a|$ here considered, the relative positions of $a^-_n$, $a^+_n$ and $a^*_n$ are linked via universal scaling relations[13]. We summarize the most relevant of such relations in Fig. 1a. Note in particular how, for $a>0$, the scaling factor between recombination minima and atom-dimer resonances is close to $\exp(\pi/2s_0) \sim 4.7$. However, the absolute location of these features depends on an unknown phase arising from non universal, short-range details of the interaction potential. The possibility of predicting such phase in real systems is still under debate[20,21]. Finally, this scenario is further enriched by the presence of two universal four-body bound states for each Efimov trimer (not shown in Fig. 1a). Such states are predicted to hit the atomic threshold for $a<0$ at positions determined by universal scaling relations[22], $a_{4,1,n}^- \sim 0.43\, a_n^-$ and $a_{4,2,n}^- \sim 0.9\, a_n^-$. We point out that in a system with finite $r_0$ the short range details of the interaction are expected to cause deviations from the universal behaviour described above[21], which become increasingly large for decreasing binding energy.

**Three-body recombination in ultracold potassium**

Recent studies on ultracold $^{133}$Cs atoms provided evidence of one three-body recombination maximum[9] and the two associated four-body resonances[22,23] for $a<0$, and of one atom-dimer resonance[10] for $a>0$ related to Efimov trimers. Unfortunately, those features are not directly connected through a universal region with $|a|=\infty$ as in the scenario above, and no information on Efimov's scaling relations can be easily extracted. We here study an ultracold gas of bosonic $^{39}$K, which combines a small background scattering length with broad magnetic Feshbach resonances[24] that allow accurate tuning of $a$ over several orders of magnitude[25,26] across a universal region. We characterized the system by studying the variation of three-body recombination processes versus $a$, and we discovered that the study of this observable alone can give informations also about the location of the atom-dimer resonances.

In the experiment, an ultracold sample, composed initially by $N_0$ atoms, was held in an optical trap and the atom number $N(t)$ was recorded after a variable interaction time $t$, over the broadest range of values of $a$ we could access. The sample was either a Bose-Einstein condensate (BEC) or a thermal cloud just above the critical temperature for condensation, to avoid as much as possible thermal effects which could mask the presence of Efimov states[12]. The recombination events were detected as a loss of atoms from the trap: the outcoming atom and dimer have a kinetic energy of the order of the binding energy of the dimer, which is typically enough for them to leave the trap. The local rate equation describing the losses is:

$$\dot{n}(t) = -n_l \alpha\, n^3(t) = -K_3\, n^3(t) \qquad (1)$$

$n$ being the atomic density, $\alpha$ the rate for three-body recombination and $n_l$ the number of atoms lost in each collision process, normally equal to three. The product $n_l\alpha$ is defined as the three-body loss coefficient $K_3$. In general, heating effects associated to the recombination processes must be taken into account by introducing an additional equation which describes the evolution of the temperature $T(t)$ of the system[18]. From the integration of these two equations, an analytic relation between $K_3$ and the atom number and temperature of the sample was obtained, allowing us to derive the $K_3$ coefficient from the experimental data (See Methods).

Using this technique we were able to investigate the regions $[-300;-8000]\, a_0$ and $[+20;+25000]\, a_0$ on the two sides of the broad Feshbach resonance in the absolute ground state[24] F=1, $m_F$=1. Comparison of the measured $K_3$ coefficient, shown in Fig. 1c, with Fig. 1a-b, suggests the presence of two different trimer states that are very close to the Efimov scenario depicted in Fig. 1a. The two maxima for $a<0$ can be attributed to a trimer state (the one at larger $|a|$) and to one of the associated four-body states. In fact, at $a_{4,1,n}^-$ and $a_{4,2,n}^-$ the enhancement of the four-body recombination is hardly distinguishable[22,23] from an enhancement of $K_3$. The two minima at $a>0$ are instead very likely associated with the trimer state above and with the next trimer state in the ladder. Finally, the two maxima we observe for $a>0$, that are not expected to appear in the three-atom recombination, are very close to the locations one would expect for the two related atom-dimer resonances. At such resonances, one can actually think of a simple second-order effect that can lead to an increase of the number $n_l$ of atoms lost in each collision, thus effectively increasing the loss coefficient $K_3$. The idea is sketched in Fig. 2a: as the atom-dimer cross section resonantly increases at $a^*_n$, an energetic dimer created in a primary recombination process can repeatedly collide with the ultracold atoms before leaving the trap. If the binding energy of the dimer (and therefore its initial kinetic energy) is large compared to the trap depth, its collision partners might gain enough kinetic energy to escape, leading to an effective $n_l>3$. As we will discuss later, this intuition is quantitatively confirmed by a simple model of the recombination dynamics.

**Fitting the experiment with the universal theory**

A comparison of Fig. 1a-b and Fig.1c already indicates that the observed anomalies in $K_3$ feature a scaling close to the geometric one expected for an Efimov spectrum in the universal regime. In order to quantitatively compare the behaviour of our system with the ideal case considered by theory, we first analyze each anomaly independently in the framework of the UT. This predicts a $K_3(a)$ for zero collision energy of the form[13]:

$$K_3(a) \approx \begin{cases} \dfrac{n_l \hbar a^4}{M}[67.12\ e^{-2\eta}(\sin^2(s_0 \ln(\dfrac{a}{a^+})) + \\ \qquad\qquad + \sinh^2(\eta)) + 16.84\,(1-e^{-4\eta})] & \text{if } a>0 \\ \dfrac{n_l \hbar a^4}{M}\dfrac{4590 \sinh(2\eta)}{\sin^2(s_0 \ln(\dfrac{a}{a^-})) + \sinh^2(\eta)} & \text{if } a<0 \end{cases} \quad (2)$$

We locally fitted the experimental data with equation (2), with $a^+$, $a^-$ and $\eta$ as fit parameters. The latter is an inelasticity parameter that describes the decay probability of the three-boson system onto deeply-bound two-body states.

Let us start from the loss minima at $a>0$. We have fitted each feature separately, finding $a^+_1=224(7)\ a_0$ with $\eta=0.043(11)$, and $a^+_2=5650(900)\ a_0$ with $\eta=0.030(17)$. The ratio of the two locations $a^+_2/a^+_1=25(4)$ is, within the experimental error, consistent with, although nominally larger than the UT prediction $\exp(\pi/s_0)\sim 22.7$.

On the other side of the Feshbach resonance, for $a<0$, the experimental $K_3$ data are only partially described by the UT behaviour of equation (2), suggesting the presence of four-body physics. The broad deviation from the $a^4$ trend exhibits two maxima centered around $-1500\ a_0$ and around $-650\ a_0$, respectively. In light of the universal connection of three- and four-body states for identical bosons[22], we interpret the maximum at $-1500\ a_0$ as a three-body Efimov resonance, and the other one as due to the associated four-body resonance[22,23]. The locations of the two features are indeed approximately in the proper ratio of 0.43 indicated by theory[22]. A fit of the main maximum with equation (2) gives $a^-_1=-1500(90)\ a_0$, and $\eta=0.14(2)$. The latter is unexpectedly larger than what is measured for $a>0$, considering that $\eta$ is expected to be constant all over a universal Efimov spectrum[13]. This might be explained by the predicted second four-body resonance at $0.90\ a^-_1$ that broadens the three-body resonance. In fact, a close inspection of the region around $-1500\ a_0$ reveals a doublet structure of the $K_3$ maximum that might be compatible with two partially resolved resonances, similarly to what observed in Ref. 23.

**Secondary losses at atom-dimer resonances**

In order to justify an increase of the three-body losses at $a^*_n$ we have constructed a simple model, similar to the one developed in Ref. 27 for studying an atomic BEC in a collisionally opaque regime. Our model computes the number of atoms $n_l$ lost from the trap via additional scattering events with a dimer created in a primary recombination process of the kind $b+b+b\to b_2^*+b$, in presence of a resonant atom-dimer scattering and a negligible atom-atom interaction. To do this the model takes into account all possible sequences of elastic and inelastic scattering events between a dimer and the surroundings atoms, which might induce the loss of the collision partner. The collision probabilities are evaluated in terms of the elastic atom-dimer cross section and of the relaxation rate for processes of the kind $b_2^*+b\to b_2+b$, for which analytic expressions are available in the low energy limit[13]. Depending on the interplay between elastic and inelastic events, the process will stop either because the kinetic energy of the dimer lowers down to values of the order of the trap depth, or because the dimer undergoes inelastic decay and is expelled from the trap with no further collisions. Note that the maximum number of secondary losses, $n_l-3$, strongly depends on the value of $E_b$. In fact, by moving towards larger $a$ values, the released energy lowers and the resonant increase of $n_l$ weakens until it disappears as $E_b$ becomes comparable to the trap depth (See Supplementary information for more details).

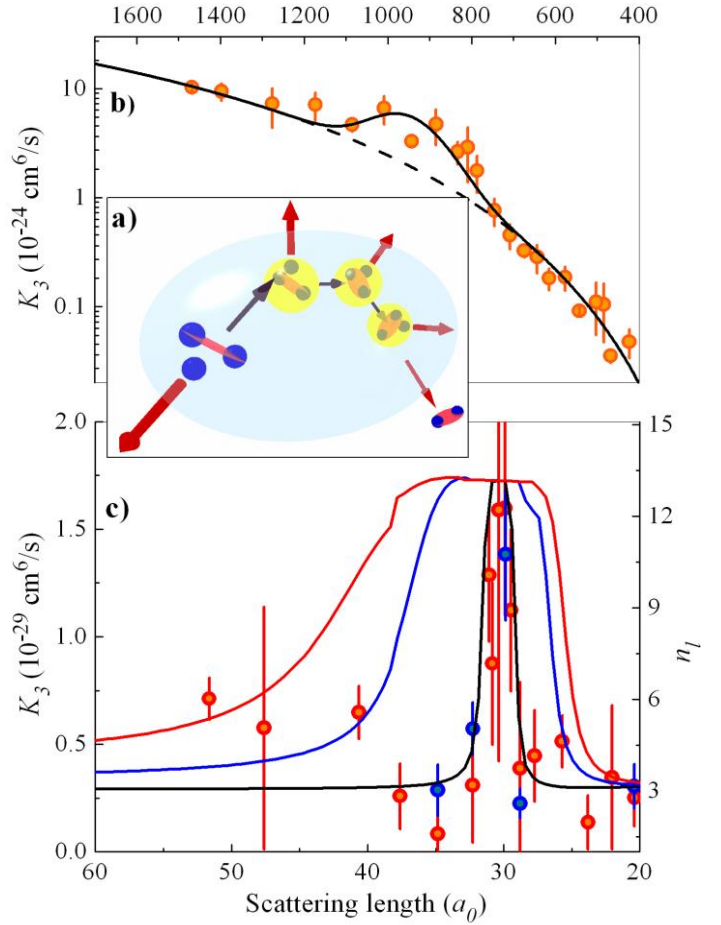

**Figure 2| Atom-dimer resonances at $a>0$. a,** Cartoon of the secondary loss mechanism at an atom-dimer resonance: a primary recombination event causes effectively $n_l=6$ atoms to be lost from the trap through secondary atom-dimer collisions (yellow). **b,** The measured $K_3$ at the second atom-dimer resonance $a^*_2$ (circles) features a deviation from the UT behaviour (dashed line) that is fitted with a Gaussian profile (solid line). **c,** Measured $K_3$ at the first atom-dimer resonance $a^*_1$ (circles). $K_3$ is extracted either by sampling $N(t)$ at $t=0$ and $t=120$ms (orange), or by sampling $N(t)$ over 8-10 times during 2.5 s of system evolution (blue). Solid lines show the calculated $K_3$ behaviour, assuming a $\sigma_{AD}$ equal to the one derived by UT divided by a factor of 1 (red), 2 (blue), or 30 (black). On the right axis, the calculated number of atoms $n_l$ lost in a single recombination event is also shown. For all data points, error bars are the root sum squared of the s.e.m. of the atom number and of the uncertainty on the trapping frequencies.

To model the loss enhancement observed at low $a$, we set $a^*_1$ to the center of the experimental feature determined by a Gaussian fit, 30.4(5) $a_0$. The results are shown in Fig. 2c. The agreement is only qualitative if we employ the low-energy expression for $\sigma_{AD}$, but it becomes also quantitative if we use a 30-times smaller $\sigma_{AD}$. We justify this mismatch by considering that our model doesn't take into account the energy dependence of the s-wave cross section: since $\sigma_{AD}$ is largest at low collision energy and since the dimer is initially out of this regime, our calculation tends to overestimate the real behaviour of the system. Our model neglects as well the effect of higher order partial waves which might contribute to the very first collisions. However, since their contribution is not in principle expected to exhibit a resonant behaviour at $a^*_n$, they might induce a shift in the background value of $n_l$, but they should not contribute to its resonant increase.

The model indicates for the feature around 30 $a_0$ that the number of particles lost in each recombination event increases from the background value $n_l$= 3 to a maximum $n_l$ ~13 due to the secondary collisions. We expect that an analogous mechanism gives rise also to the second feature peaked at $a^*_2$ (although the maximum $n_l$ is significantly reduced because of the smaller $E_b$). However, we cannot apply our model to such strongly interacting regime. In this case, a fit of the data with a Gaussian function superimposed to equation (2) sets $a^*_2$=930(40) $a_0$. Note that this secondary process is not peculiar of our system, but is in principle observable whenever the initial energy of the dimer is large enough to cause a perceptible increase of atom losses via secondary elastic collisions. This makes the effect of quite general interest for Efimov studies, since it avoids the need of studying further observables such as the atom-dimer cross-sections.

## Scaling factors and deviations from universality

The analysis performed above indicates that all features discovered in our system are locally well described by the three- and four-body theories currently available. To further discuss their global behaviour, we have summarized in Table 1 the main ratios between the positions of the observed features and compared them with the UT predictions. Note how the two recombination minima and the two atom-dimer resonances are close to the expected scaling depicted in Fig. 1a, although their ratios appear systematically larger than the universal ones. A simultaneous fit of the two minima with equation (2), employing the Efimov scaling factor $s_0$ as a fit parameter actually gives $s_0$=0.956(18), confirming a larger spacing for the $a>0$ features. For $a<0$ the position of the only three-body feature we observe at $a^-_1$ deviates substantially from the expected ratio with the position of the other extreme of the trimer state at $a^*_1$. It is however closer to the UT ratio with the extreme of the next trimer state at $a^*_2$. The deviations of the spectrum observed in the experiment from the UT predictions can be actually attributed to the finite $r_0$ of our system. As we already pointed out, it is indeed known[21,28,29] that when $|a|$ becomes on the order of $r_0$, which in turn is on the order of the Van der Waals radius[24] of our system, 64 $a_0$, the UT predictions start to be affected by the short-range details of the potential. In particular, recent theoretical studies[21,28,29] indicate that a finite range of the interaction potential can lead to a larger (smaller) binding energy of the lowest Efimov states for $a>0$ ($a<0$), causing a correspondingly larger (smaller) scaling factor. Our observations seem to fit these expectations: the larger ratios for $a>0$ in Table 1 are apparent and the shift in $a^-_1$ might be due to a smaller scaling factor for $a<0$. Note how the last column in Table 1 indicates a clear trend for the deviation of the experimental data from the UT prediction. The deviation decreases rapidly as $|a|$ increases beyond the Van der Waals radius, indicating that the system is getting progressively deeper into the universal regime.

|  | Ratio | Exp | UT | (Exp-UT)/UT (%) |
|---|---|---|---|---|
| $a>0$ | $a^+_1/a^*_1$ | 7.4(3) | 4.5 | +64(7) |
|  | $a^*_2/a^*_1$ | 30.6(14) | 22.7 | +35(6) |
|  | $a^+_2/a^*_2$ | 6(1) | 4.5 | +33(22) |
|  | $a^+_2/a^+_1$ | 25(4) | 22.7 | +10(18) |
| $a<0$ | $a^*_1/a^-_1$ | -0.020(1) | -1/21.3 | -57(3) |
|  | $a^*_2/a^-_1$ | -0.62(5) | -22.7/21.3 | -42(4) |

**Table 1| Experimental vs Universal ratios.** Comparison of some ratios between the scattering length values of the observed features (Exp) and the ones expected in the universal regime (UT). Note how the relative disagreement between experiment and universal trend, shown in the last column, decreases as $|a|$ increases.

Further comparison with advanced theories might clarify whether these deviations can still be described in the frame of a universal model for finite range interactions[28-30]. Finally, it is interesting to compare our observations with the recent theoretical prediction of Ref. 20, in which the three-body short range phase was derived on the base of the sole knowledge of the two-body physics[24]. In particular, the theoretical predictions for $a>0$ show a remarkable agreement with most of our observations: $a^*_1$=30.7, $a^+_1$=212 $a_0$, $a^*_2$=810 $a_0$ and $a^+_2$=3850 $a_0$. This keeps open the question about the predictive power of low-energy models for the location of three-body features[21], which might find an answer with further comparison with experimental data.

## Outlook

In conclusion, we have detected two consecutive Efimov trimer states and of at least one four-body state in a system of bosonic potassium atoms with tunable scattering length. The spectrum we have observed features a scaling that seems to tend to the universal geometric one as the scattering length increases well beyond the two-body characteristic radius. Our system realizes a test bed for advanced few-body theories that include short-range effects, and might provide further insight on the three- and four-body physics in future experiments. For instance, one could compare distinct Efimov spectra on the same atomic system by employing different Feshbach resonances[24]. This might help in establishing a possible link between two-body parameters and the short-range three-body phase, and could clarify to what extent the presence of a finite range of the potential can still lead to universal effects[28-30]. Finally, a direct measurement of the binding energy of the trimers across the resonance, a task that is complicated by the very short lifetime of such states, would provide an even more stringent test of the universal three-body binding laws.

## Methods

**Experimental procedure.** The scattering length of $^{39}$K atoms in their ground state can be varied across a Feshbach resonance by varying a homogeneous magnetic field $B$, with a law that can be rather well parameterized as $a(B)/a_0 = a_{bg}(1-\Delta/(B-B_0))$. Here $a_{bg}$=-29.0(3) $a_0$ is from Ref. 24; $B_0$=402.50(3) G and $\Delta$=−52.1(1) G are derived from Ref. 24 and from recent measurements (See Supplementary information). Our accuracy on $B$ is 30mG. Note also that the absence in our system of weaker Feshbach resonances close to the broad one allows us to directly ascribe all anomalies in the $\alpha$ coefficient to three- and four-body physics. We started all measurements with an almost pure BEC of about $1\times10^5$ atoms with $a$=180 $a_0$, loaded in an optical trap with mean frequency $\nu$=75(5) Hz, following a procedure already described elsewhere[25]. We used slightly different procedures to investigate the various regions of $a$ values shown in Fig.1. To investigate the $a$>0 region, the field was adiabatically ramped to the value of interest and kept constant for a variable holding time as long as 2.5s. After this time the thermal component of the BEC remained negligible for $a$ up to 300 $a_0$. For 300 $a_0$ < $a$ < 600 $a_0$ a partially condensed sample was left. For even larger $a$ the system rapidly turned into a thermal gas at $T$~100 nK. To investigate the $a$<0 region, $a$ was rapidly set to -110 $a_0$, where the system collapses and re-equilibrates to a thermal gas at about the critical temperature for condensation. The $B$ field was then ramped to the value of interest, and held there for a measurement time as long as 2.5 s. More detailed information on the actual experimental parameters can be found on the online Supplementary Information.

**Determination of $K_3$.** To determine the $K_3$ coefficient from the measured atom number $N(t)$ we integrated equation (1), taking into account also the heating effect associated to the atom losses. In the case of a thermal sample the losses remove from the sample preferentially the coldest atoms, leading to an increase of the temperature that can be easily taken into account[18]. In the case of a quasi-pure BEC, the total energy per particle is uniform and therefore losses will increase the temperature of the system mainly by reducing the condensed fraction. This is however a minor effect if losses are not too large. Finally, since a finite temperature $T$ limits the $K_3$ in equation (2) to a maximum value $K_{3max} \propto 1/T^2$ (the so-called unitary limit)[12], we include this effect by employing as a fitting function[31] $(1/K_3 + 1/K_{3max})^{-1}$ for the data at $a$>0, where the saturation of $K_3$ due to thermal effects is evident at very large values of $a$.

In all these cases, it is possible to find an analytic expression for the time evolution of the atom number (at least in an implicit form) which allows us to extract an expression for the bare $K_3$ from the experimental data.

A summary of the expressions used to determine $K_3$ from the data can be found in the Supplementary information. A large fraction of the $K_3$ data derives from measurement of $N$ at two times only ($t$=0 and $t$=$t_H$), a procedure that we have experimentally verified to give the same result and a comparable uncertainty as measurement of the full time-dependence of $N$ (see for instance Fig.2c). In both cases, error bars are evaluated as the root sum squared of the standard error of the mean value (s.e.m.), given by the fit and of $\delta\nu\times(\partial K_3/\partial\nu)$, $\delta\nu$ being the mean trap frequency uncertainty.

**Acknowledgements** We thank P. Massignan and H. Stoof for motivating this work and for discussions. We acknowledge also B. D. Esry, J. P. D'Incao, R. Grimm and co-workers, and A. Simoni for discussions. We are grateful to S. Bartalini and G. Thalhammer for technical support. This work has been supported by CNR within the EUROCORES Programme EuroQUAM of the European Science Foundation, by the European Research Council through a Starting Grant, and by the Italian MIUR (PRIN2006 programme).

**Additional information** Correspondence and requests for materials should be addressed to M. Z. (zaccanti@lens.unifi.it).

**Competing financial interests** The authors declare no financial interests.

1. Burnett, K., Julienne, P. S., Lett, P. D., Tiesinga, E., and Williams, C. J., Quantum encounters of the cold kind, *Nature* **416**, 225-232 (2002).
2. Köhler, T., Goral, K., and Julienne, P., Production of cold molecules via magnetically tunable Feshbach resonances, *Rev. Mod. Phys.* **78**, 1311-1361 (2006).
3. Jensen, A. S., Riisager, K., Fedorov, D. V. & Garrido, E. Structure and reactions of quantum halos. *Rev. Mod. Phys.* **76**, 215-261 (2004).
4. Efimov V., Weakly-bound states of three resonantly-interacting particles. *Sov. J. Nucl. Phys.* **12**, 589-595 (1971).
5. Efimov, V. Energy levels arising from resonant two-body forces in a three-body system, *Phys. Lett. B* **33**, 563-564 (1970).
6. Thomas, L. H., The interaction between a neutron and a proton in the structure of $H^3$, *Phys. Rev.* **47**, 903-909 (1935).
7. Efimov, V., Energy levels of three resonantly interacting particles, *Nucl. Phys. A*, **210**, 157-188 (1973).
8. Brühl, R. et al., Matter wave diffraction from an inclined transmission grating: Searching for the elusive $^4$He trimer Efimov state. *Phys. Rev. Lett.* **95**, 063002 (2005).
9. Kraemer, T., et al., Evidence for Efimov quantum states in an ultracold gas of caesium atoms, *Nature* **440**, 315-318 (2006).

10. Knoop, S. et al., Observation of an Efimov-like resonance in ultracold atom-dimer scattering, *Nat. Phys.* **5**, 227 – 230 (2009).

11. Tiesinga, E., Verhaar, B. J. & Stoof, H. T. C. Threshold and resonance phenomena in ultracold ground-state collisions. *Phys. Rev. A* **47**, 4114-4122 (1993).

12. D'Incao, J. P., Suno, H. & Esry, B. D. Limits on universality in ultracold three boson recombination. *Phys. Rev. Lett.* **93**, 123201 (2004).

13. Braaten, E. & Hammer, H.-W. Efimov physics in cold atoms. *Annals of Physics* **322**, 120-163 (2007).

14. Bedaque, P. F., Braaten, E. & Hammer, H. W. Three body recombination in Bose gases with large scattering length. *Phys. Rev. Lett.* **85**, 908-911 (2000).

15. Petrov, D. S. Three-boson problem near a narrow Feshbach resonance, *Phys. Rev. Lett.* **93**, 143201 (2004).

16. Esry, B. D., Greene, C. H. & Burke, J. P. Jr Recombination of three atoms in the ultracold limit. *Phys. Rev. Lett.* **83**, 1751-1754 (1999).

17. Fedichev, P. O., Reynolds, M. W & Shlyapnikov, G. V. Three-Body Recombination of Ultracold Atoms to a Weakly Bound *s* Level. *Phys. Rev. Lett.* **77**, 2921-2924 (1996).

18. Weber, T, et al., Three-body recombination at large scattering lengths in an ultracold atomic gas. *Phys. Rev. Lett.* **91**, 123201 (2004).

19. Nielsen, E. & Macek, J. H., Low-Energy Recombination of Identical Bosons by Three-Body Collisions. *Phys. Rev. Lett.* **83**, 1566-1569 (1999).

20. Massignan, P. & Stoof, H., Efimov states near a Feshbach resonance. *Phys. Rev. A* **78**, 030701(R) (2008).

21. D'Incao, J.P., Greene, C.H. & Esry, B. D., The short-range three-body phase and other issues impacting the observation of Efimov physics in ultracold quantum gases. *J. Phys. B: At. Mol. Opt. Phys.* **4**, 044016 (2009).

22. von Stecher, J., D'Incao, J. P. & Greene, C. H., Four-body legacy of the Efimov effect. arXiv:0810.3876.

23. F. Ferlaino, S. Knoop, M. Berninger, W. Harm, J. P. D'Incao, H.-C. Nägerl, and R. Grimm, Evidence for Universal Four-Body States Tied to an Efimov Trimer, *Phys. Rev. Lett.* **102**, 140401 (2009).

24. D'Errico, C., et al., Feshbach resonances in $^{39}$K, *New J. Phys.* **9**, 223 (2007).

25. Roati, G. et al., $^{39}$K Bose-Einstein condensate with tunable interactions. *Phys. Rev. Lett.* **99**, 010403 (2007).

26. Fattori, M., et al., Magnetic Dipolar Interaction in a Bose-Einstein Condensate Atomic Interferometer. *Phys. Rev. Lett.* **101**, 190405 (2008).

27. Schuster, J. et al., Avalanches in a Bose-Einstein condensate. *Phys. Rev. Lett.* **87**, 170404 (2001).

28. Platter, L., Ji, C. & Phillips D. R. Range corrections to three-body observables near a Feshbach resonance. *Phys. Rev. A* **79**, 022702 (2009).

29. Thøgersen, M., Fedorov, D. V., and Jensen, A. S., Universal properties of Efimov physics beyond the scattering length approximation, *Phys. Rev. A* **78**, 020501 (2008).

30. H.-W. Hammer, Timo A. Lähde, and L. Platter, Effective-range corrections to three-body recombination for atoms with large scattering length, *Phys. Rev. A* **75**, 032715 (2007).

31. Greene, C. H., Esry, B. D., and Suno, A., A revised formula for 3-body recombination that cannot exceed the unitarity limit. *Nuclear Physics A* **737**, 119-124 (2004).

**Supplementary Information**

**Model for atom losses at dimer-trimer crossings**

We developed a simple model to estimate the number of atoms $n_l$ lost per recombination event at the dimer-trimer crossings, where the atom-dimer scattering features a resonant behavior. We assume herein that dimers are formed via three body recombination processes at the center of the atomic cloud, and that atom-atom interactions are negligible with respect to the atom-dimer ones. The initial kinetic energy of the dimer is $E_b/3$, with $E_b$ the dimer binding energy. We define the following quantities: 1) $p_e$ is the probability for a dimer to undergo an elastic collision with an atom before leaving the trap. This is given by[27] $p_e = 1-exp(-\langle n\,l\rangle \sigma_{AD})$, where $\langle n\,l\rangle$ is the mean column atomic density and $\sigma_{AD}$ the atom-dimer elastic cross section. 2) $p_i(0)$ is the probability for a dimer to undergo an inelastic collision before the first elastic collision takes place. This probability can be estimated evaluating the relative number of dimers lost at the time of the first elastic collision $\tau_0=1/(nv_0\,\sigma_{AD})$, $v_0 = (E_b/3M)^{1/2}$ being the dimer initial velocity. This can be obtained from the evolution of the dimers undergoing inelastic collisions with atoms, described by the following rate equation:

$$\dot{n}_D(t) = -\beta_{AD} n_A(t) n_D(t) \quad \text{(SI1)}$$

where $\beta_{AD}$ is the event rate for dimers undergoing inelastic collisions with atoms. In the limit of few three-body processes, the atom density greatly exceeds the one of dimers, and equation (SI1) can be integrated assuming $n_A$ to be constant. This immediately leads to an exponential decay of the dimer density, $n_D(t)=n_D(0)\,exp(-\beta_{AD} n_A t)$, and the probability of having an inelastic decay after time $\tau_0$ can be estimated as $p_i(0)=1-exp(-\beta_{AD}/v_0\sigma_{AD})$. Both for $\sigma_{AD}$ and $\beta_{AD}$ analytic expressions are available in the low energy limit[13]. 3) Finally, we define the probability for inelastic decay of the dimer after k-1 elastic collisions as $p_i(k)= p_i(t=\tau_0+\tau_1+...+\tau_k)$. This is easily evaluated by considering that the mean energy of the dimer after an elastic collision with an atom at rest is 5/9 of the initial one, so that

$$p_i(k) = p_i(0)\sum_{i=0}^{k}(9/5)^{i/2} \quad \text{(SI2)}$$

Each elastic collision causes the loss of an atom until the dimer energy $(E_b/3)(5/9)^{kM}$ becomes of the order of the trap depth $E_T$, for a certain $k_M$; then, the dimer remains trapped, until it undergoes an inelastic collision. The inelastic decay will surely take place for a certain $k_S$ such that $p_i(k_S) \geq 1$. Depending on the interplay between elastic and inelastic atom-dimer collisions, $k_S$ can be either larger or smaller than $k_M$. We define $K_C = Min\{k_M, k_S\}$. Taking into account all possible sequences of scattering events, we deduce the following expression for $n_l$:

The expression is normalized in order to guarantee that the dimer has an overall probability of escaping the trap equal to one: this also ensures that the maximum atom number lost via secondary collisions is less or equal to $3+k_M+1$. Then, all parameters in (SI3) can be easily calculated from analytical expressions in the zero collision energy regime[13]. Since $\sigma_{AD}$ is largest in the s-wave regime and since the dimer is initially out of this regime, the evaluation of $n_l$ in (SI3) is an overestimation of the real behaviour of the system. However, the model can be easily extended by inclusion of finite collision energy effects, provided that the elastic and inelastic atom-dimer scattering cross sections are evaluated for finite collision energy and higher partial waves.

The theoretical curves in Fig. 2c were evaluated for a BEC of $7.5\times 10^4$ atoms loaded in a trapping potential with frequencies (87,141,141) Hz and a depth of 1.5 $\mu$K. In the analytic formulae for elastic cross section and inelastic decay rate we set $a^*=30.4\,a_0$ and $\eta=0.01$. In order to evaluate $k_M$ for different scattering length values, the binding energy of the Feshbach molecule as a function of the magnetic field is needed: in the region of interest, its behaviour is linear with a magnetic moment of [24] -2.1 MHz/Gauss, that sets $k_M \sim 10$.

**Determination of $K_3$ from $N(t)$**

In the case of a thermal gas, deriving the time evolution N(t) of the atom number by integrating the local rate equation

$$\dot{n}(t) = -n_l \alpha\, n^3(t) = -K_3 n^3(t) \quad \text{(SI4)}$$

is a non trivial problem: recombination events not only modify the density distribution of the sample, but also tend to increase the mean particle energy in the system by preferentially removing less energetic atoms residing in the denser part of the cloud. The heating of the gas must be taken into account introducing a coupled equation for the temperature[18]. In general, for $a > 0$, if the energy $E_b$ of the molecular state is comparable with the trap depth, the products of the three-body recombination process can also remain trapped. While the dimer is quickly lost after a successive inelastic decay onto a deeply bound state, the atom can stay in the trap with a kinetic energy of $2E_b/3$, inducing an extra heating. In our case, this mechanism has an effect only for very high values of $a$, and therefore we can neglect it. The coupled equations we have to solve are therefore:

$$\frac{dN}{dt} = -\beta K_3 \frac{N^3}{T^3}$$

$$\frac{dT}{dt} = \frac{\beta K_3}{3}\frac{N^2}{T^2} \quad \text{(SI5)}$$

$$n_l = 3 + \frac{\sum_{k=0}^{Kc} p_e^k\, k(1-p_i(k+1))(1-p_e) + \sum_{k=0}^{Kc}(k+1)p_e^k\, p_i(k+1) + \sum_{k=1+k_M}^{k_S}(k_M+1)p_e^k\, p_i(k+1)}{\sum_{k=0}^{Kc} p_e^k (1-p_i(k+1))(1-p_e) + \sum_{k=0}^{Kc} p_e^k\, p_i(k+1) + \sum_{k=1+k_M}^{k_S} p_e^k\, p_i(k+1)} \quad \text{(SI3)}$$

where $\beta = \left(\dfrac{M(2\pi\nu)^2}{2\pi k_B}\right)^3 \dfrac{1}{\sqrt{27}}$, with M the mass of the particles and $\nu$ the mean trap frequency. Their solution is:

$$N(t) = \dfrac{N_0}{(1 + (3\beta N_0^2 t / T_0^3) K_3)^{1/3}} \quad \text{(SI6)}$$

$$T(t) = T_0 (1 + (3\beta N_0^2 t / T_0^3) K_3)^{1/9}$$

$N_0$ and $T_0$ being the initial atom number and temperature.

The case of a pure BEC is much simpler to handle, since the total energy per particle is uniform over its density distribution. Nevertheless, at finite temperature the losses will induce a decrease of the condensed fraction and therefore a heating; however, this effect remains negligible if the losses in the system are not too large. Therefore, one can integrate equation (SI4) assuming at each time that the BEC has its steady density distribution. In this case the atom number evolves as:

$$N(t) = \dfrac{N_0}{(1 + (4/5 A N_0^{4/5} t / a^{6/5}) K_3)^{4/5}} \quad \text{(SI7)}$$

with $A = \dfrac{1}{3!} \dfrac{5^{4/5}}{56\, \pi^2 3^{1/5}} \left(\dfrac{2\pi\nu M}{\hbar}\right)^{12/5}$, and $a$ being the scattering length. In both cases, one can use equation (SI6) and (SI7) as fitting functions for the time evolution of the atom number of the system and derive $K_3$. Otherwise, one can measure the initial atom number and temperature $N_0$ and $T_0$, and the $N(t_H)$ after some fixed holding time $t_H$ and determine the $K_3$ by inverting equations (SI6) and (SI7).

In the case of very high $a$, the three-body loss coefficient $K_3$ saturates to a maximum value usually called *unitary limit*, which depends by the temperature $T$ of the system as[12]:

$$K_{3\max}(T) = \dfrac{128 \varepsilon \pi^2 \hbar^5}{M^3 T^2} \equiv \dfrac{\Delta}{T^2} \quad \text{(SI8)}$$

where $\varepsilon$ is a multiplicative constant expected to be[12] of the order of 0.1. As the system evolves, the temperature increases and consequently the unitary limit decreases. Such effect can be directly taken into account[31] introducing in equation (SI5) a total three-body losses coefficient defined as:

$$K_{3TOT}(T) = \dfrac{K_{3\max}(T) K_3}{K_{3\max}(T) + K_3} \quad \text{(SI9)}$$

In this case, the solution for $N$ and $T$ can be found only in an implicit form:

$$T(N) = T_0 \left(\dfrac{N_0}{N}\right)^{1/3}$$

$$t = \dfrac{3}{\beta K_3 \Delta N_0^2 T_0^6} \left\{ \dfrac{\Delta T_0^9}{9} \left[\left(\dfrac{N_0}{N(t)}\right)^3 - 1\right] + \dfrac{K_3 T_0^{11/3}}{11} \left[\left(\dfrac{N_0}{N(t)}\right)^{11/3} - 1\right] \right\} \quad \text{(SI10)}$$

However, it is still possible to invert it in order to extract an expression for the bare $K_3$. Note that in the limit $\Delta/T^2 \gg K_3$ the equation (SI10) becomes equation (SI6). In the evaluation of $K_3$ we have set $\varepsilon \sim 0.2$ to get the best agreement with theory.

The regions around $a_1^*$ and $a_1^+$ were fitted with equation (SI7), for $a < 600\, a_0$. Above this value, and for the data in the regions around $a_2^*$, $a_2^+$ and $a_{4,1}^-$ we employed equation (SI6). Around $a_1^-$ we used equation (SI10) in order to extract the bare $K_3$. Note that the evaluation with equation (10) almost coincides with the one derived using equation (SI6) for data at $|a| < 3000\, a_0$.

In fitting the behaviour of a thermal gas case, we completely neglected mean-field effects arising for large $|a|$ values, which might be expected to modify the density distribution of the sample and therefore the evaluation in (SI5). We estimate their relevance by comparing the thermal energy $k_B T$ with the mean-field energy evaluated as

$$E_{MF} = \dfrac{4\pi\hbar^2}{M} \dfrac{a}{\sqrt{1 + (ka)^2}} n_M,$$

where $k = \sqrt{3M k_B T}/\hbar$ is the thermal wave-vector, and $n_M$ is the mean-density of a Maxwell-Boltzmann distribution. For our parameters, $E_{MF}$ never exceeds $0.21 \times k_B T$. Consequently, we estimate that the error on $K_3$ due to neglecting mean-field effects in the evaluation is always smaller than 50%. We point out that an error in the determination of the atom number from absorption imaging as large as 30% can lead to a systematic shift of the $K_3$ as large as 200% not considered in the error bars in Figures 1 and 2.

The mean trapping frequency $\nu$, the initial atom number $N_0$ and temperature $T_0$, and maximum holding time $t_H$ vary depending on the range of values of $a$ we investigated. $a_1^*$ was studied with a pure BEC of of $N_0 = 7.5(5) \times 10^4$ atoms stored for up to $t_H = 2.5$ s in a $\nu = 120$ Hz trap. For $a_1^+$ we employed a pure BEC of $N_0 = 1.3(5) \times 10^5$ atoms stored for up to $t_H = 1$ s in the a $\nu = 75(5)$ Hz trap. The region around $a_2^*$ was studied with a BEC of of $N_0 = 7.0(5) \times 10^4$ atoms stored for up to $t_H = 170$ ms in the a $\nu = 75(5)$ Hz trap: for data above $a > 600\, a_0$ a ~100 nK thermal cloud was left after few ms. For $a_2^+$ we employed a thermal cloud of $N_0 = 2.1(1) \times 10^4$ atoms at $T_0 = 80(5)$ nK stored for $t_H = 65$ ms in a $\nu = 67(4)$ Hz trap. The $a < 0$ side of the resonance was investigated with thermal samples just above the critical temperature for condensation. For the region around $a_1^-$ we employed a thermal cloud of $N_0 = 2.0(1) \times 10^4$ atoms at $T_0 = 80(5)$ nK stored for $t_H = 45$ ms in a $\nu = 67(4)$ Hz trap. For the four-body resonance region around $-650\, a_0$ we employed a thermal gas of $N_0 = 1.7(1) \times 10^4$ atoms at $T_0 = 42(3)$ nK stored for $t_H = 100$ ms in a $\nu = 36(2)$ Hz trap.

**Determination of the center of the Feshbach resonance**

The Feshbach resonance we employed in this work for tuning the scattering length in our system has been experimentally observed for the first time in 2007 by our group by studying both atom losses and molecule formation. This was reported in Ref. 24, together with the observation of seven other resonances in different internal states of $^{39}$K. This experimental characterization allowed us to optimize a quantum collisional model, which we employed to predict the magnetic-field dependence of scattering lengths and of near-threshold molecular levels of potassium isotopes. In Ref 24, the center of the resonance $B_0$ was found to be at 402.4(2) Gauss. The uncertainty on $B_0$ is responsible for the most important systematic error in the determination of the scattering length as one moves into the strongly interacting regime. We reduced this uncertainty down to 30 mGauss by performing an additional, more sensitive, investigation of the resonance region. This was done by studying the effects of the resonantly large interaction in a very dilute sample during ballistic expansion from the trap, a technique that allows us to get very narrow resonant features. We initially prepared 60(5) $\times 10^3$ potassium atoms just above the condensation temperature in an optical trap with a mean frequency of 75(5) Hz, by following a procedure similar to the one described in the Methods. The magnetic field was set at 396 Gauss, corresponding to a scattering length of 200 $a_0$. We then switched off the trapping potential and let the system freely expand. After about 10 ms of ballistic expansion, in a time of about 1 ms we set the magnetic field to a final variable value $B_f$. We let the system expand within $B_f$ for other 10 ms, and we then switched off the magnetic field. After other 15 ms we took an absorption image of the atomic cloud. The presence of the resonance was mostly revealed from an enhancement of the size of the atomic cloud shown in Fig. SI 1, for which we have the following explanation. As the scattering length is increased towards large and positive values the mean field energy of the system increases. This leads to a faster expansion of the system and therefore to a larger size after expansion. As soon as the scattering length is tuned across the resonance center to very large, negative values, the system undergoes mean field collapse, accompanied also by an enhancement of three-body processes which heats up the sample. Again, this turns into an enhancement of the size of the cloud measured at long expansion times. As the scattering length is tuned towards smaller $|a|$ the mean field effects rapidly decreases, leading to an almost unperturbed ballistic expansion. Since a larger size after expansion corresponds to a larger released energy, we assume the resonance center to coincide with the maximum of the cloud size. A gaussian fit of the data sets the center of the feature at 402.50 Gauss with an uncertainty of 14 mGauss. By taking into account also the short term stability of the magnetic field we have in the experiment, the overall uncertainty on the resonance location becomes about 30 mGauss.

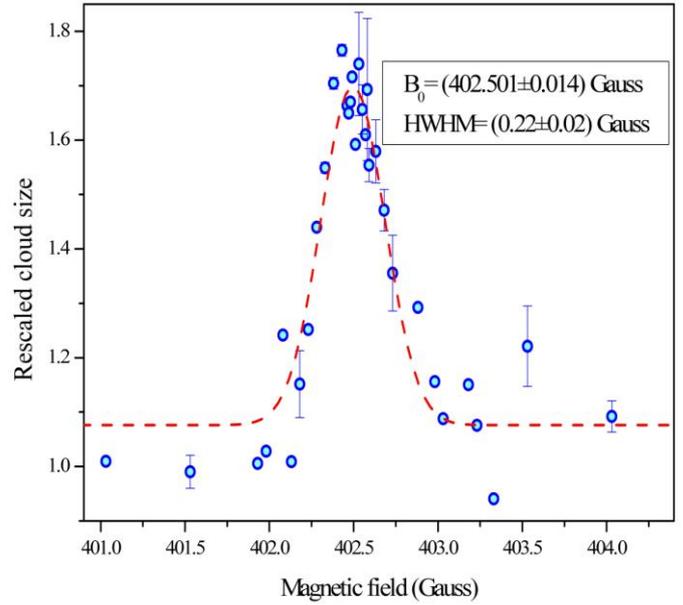

**Supplementary Figure 1| Determination of the resonance center.** The center of the resonance is revealed by studying the behaviour of a thermal cloud at the critical temperature for condensation subjected, during the expansion, to a change of the scattering length (see text). The increase of mean field energy is detected as an enhancement of the cloud radius. The relative increase of the size of the system (blue circles) is fitted with a gaussian profile (red dashed line). The resonance center is determined with an overall uncertainty of about 30 mGauss, which is the root sum squared of the uncertainty of the fit (14 mGauss) and of the stability of the magnetic field in the experiment.